\documentclass[doublecol]{epl2}
\usepackage{graphicx}
\usepackage{float}
\usepackage{amssymb}
\usepackage[utf8]{inputenc}
\usepackage[french,english]{babel}
\usepackage[intlimits]{amsmath}
\usepackage{mathtools}
\usepackage{amsfonts}
\usepackage{cases}
\usepackage{bm}%
\usepackage{hyperref}

\newcommand{\be}{\begin{align}}
\newcommand{\ee}{\end{align}}
\newcommand{\dif}{\mathrm{d}}

\newcommand{\C}{\mathbb{C}}

 \newcommand{\im}{\mathbf{i}}
  
 \newcommand{\abs}[1]{\left\vert#1\right\vert}

 \usepackage{color}
\usepackage{verbatim}

\newcommand{\red}[1]{{\color{red} #1}}

\definecolor{blue}{rgb}{0,0,1}
\definecolor{green}{rgb}{0,1,0}
\definecolor{red}{rgb}{1,0,0}
\definecolor{vio}{rgb}{1,0,1}
\definecolor{uv}{rgb}{0.5,0,0.5}
\definecolor{ama}{rgb}{0.3,0.3,0.3}

\title{Joint min-max distribution and Edwards-Anderson's order parameter of the circular $1 / f$--noise model}
\author{Xiangyu Cao \inst{1} \and Pierre Le Doussal \inst{2}}

\institute{
  \inst{1} LPTMS, CNRS, Univ. Paris-Sud, Université Paris-Saclay, 91405 Orsay, France\\
  \inst{2} CNRS-Laboratoire de Physique Théorique de l'\'Ecole Normale Supérieure, 24 rue Lhomond, 75231 Paris Cedex, France\\
}
\pacs{05.40.-a}{Fluctuation phenomena, random processes, noise, and Brownian motion}
\pacs{64.70.Q-}{Theory and modeling of the glass transition}
\abstract{
We calculate the joint min--max distribution and the Edwards-Anderson's order parameter for the circular model of $1 / f$--noise. Both quantities, as well as generalisations, are obtained exactly by combining the freezing-duality conjecture and Jack-polynomial techniques. Numerical checks come with significantly improved control of finite-size effects in the glassy phase, and the results convincingly validate the freezing-duality conjecture. Application to diffusive dynamics is discussed. We also provide a formula for the pre-factor ratio of the joint/marginal Carpentier-Le Doussal tail for minimum/maximum which applies to any logarithmic random energy model. }
\begin{document}
\maketitle
The statistical physics of a particle in logarithmically correlated random potentials was initially studied as simplified models of spin glass known as logarithmic Random Energy Models (log-REM's) \cite{derrida1980random,derrida1988polymers,carpentier2001glass}, but is now realised to be relevant to subjects ranging from multi-fractal wave-functions \cite{chamon1996localization,castillo97dirac}, extrema of 2d Gaussian Free Field (GFF) \cite{daviaud2006extremes,ding2014extreme} and 2d quantum gravity \cite{duplantier2014critical}, to the value distribution of random matrix characteristic polynomials and the Riemann zeta on the critical line \cite{fyo12zeta,fyodorov2014freezing,arguin2015maxima}. 

A key feature of log-REM's is \textit{freezing}, \textit{i.e.}, temperature--independence of the free energy density in the glassy phase. The extension of freezing to describe the free energy \textit{fluctuation} has a long history \cite{derrida1988polymers,carpentier2001glass,fyodorov08rem,fyodorov2009statistical} and was recently promoted to a rigorous stage \cite{madaule2013glassy} by using derivative multiplicative chaos \cite{duplantier2014critical,duplantier2014log}. Yet, explicit predictions (\textit{e.g.}, of free energy distribution) still require non-rigorous approaches and some ``integrability'' coming from log-gas integrals \cite{forrester2010log}, $\beta$--random matrix theory \cite{dumitriua2002matrix}, or symmetric functions \cite{macdonald1998symmetric}.

From the study of the latter, accumulating evidence supports the freezing duality conjecture (FDC), first put forward in \cite{fyodorov2009statistical}. To describe it, consider any thermodynamic observable $\mathcal{O}(\beta)$, supposed analytical in $\beta$ for $0 <\beta < \beta_c$ and \textit{analytically} continued to $\mathcal{O}_a(\beta), \beta \in (0, \infty)$. The FDC claims that, if $\mathcal{O}_a(\beta)$ is \textit{duality-invariant}, $\mathcal{O}(\beta)$ \textit{freezes}: 
\begin{align}  \mathcal{O}_a(\beta) \stackrel{\forall \beta}= \mathcal{O}_a(\beta_c^2 / \beta) \Rightarrow \mathcal{O}(\beta) \stackrel{\beta > \beta_c}= \mathcal{O}(\beta_c)\,.\label{eq:freezdual}\end{align}
$\mathcal{O}$ can be observables not yet covered by the rigorous results, \textit{e.g.} moments of the minimum position \cite{fyodorov2010freezing,fyodorov2015moments}. The theoretical understanding of 
the FDC is unsatisfactory and tests of its predictions remain limited in both quantity and quality (due to slow numerical convergence in the glassy phase).

\begin{figure}
\center \includegraphics[scale=0.35]{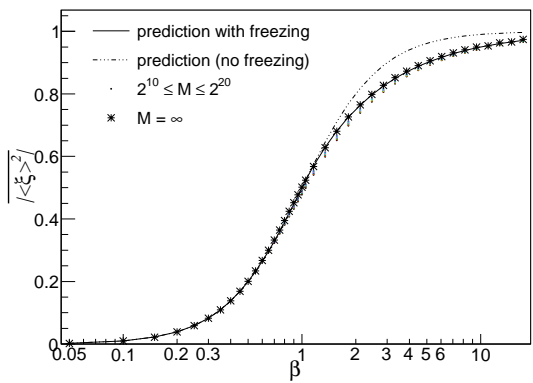}
\caption{Numerical determination of the Edwards-Anderson order parameter $\overline{\abs{\langle \xi \rangle}^2}$, as a function of inverse temperature $\beta = 1/T$.  The $M = \infty$ value is extrapolated from data $M = 2^{10}\dots 2^{20}$ using the linear Ansatz $a_0 + a_1 / \ln M$ point-wise. The result is in excellent agreement with the prediction \eqref{eq:anderfreez} based on freezing. For comparison, we also plot the prediction without freezing by analytically continuing the $\beta < 1$ expression of \eqref{eq:anderfreez}.}  \label{fig:anderson}
\end{figure}

This Letter improves significantly the situation by studying the \textit{circular model}, introduced in \cite{fyodorov08rem}, where the distribution of the minimum was calculated. We show here that there is an infinite series of duality-invariant observables, of which the simplest is the Edwards-Anderson's (EA) order parameter, fundamental in spin-glass theory \cite{edwards1975theory}. It provides one of the most accurate numerical test of the FDC (\textit{c.f.} fig. \ref{fig:anderson}).

We also calculate the\textit{ joint} min-max distribution. As application, we obtain the distribution of the \textit{span} (the min-max difference), which is the extremal width of interfaces modeled by the log-correlated field studied in experiments \cite{aarts2004direct,devilleneuve2008statistics}. Moreover, properties of opposite extrema are related to the \textit{dynamics} of an over-damped diffusive (Langevin) particle in the 1d potential. Roughly, the span is the barrier that the particle should surmount to explore the whole system, and is thus related to Arrhenius passage times and to the diffusion coefficient in the periodic potential \cite{derrida1983velocity,ledoussal1995creep,dean14diffusion}. In the log-correlated 1d case, the freezing of log-REM's is directly responsible for the freezing of dynamical exponents \cite{castillo2001freezing}. Since the opposite extrema are far apart in space and in value, they are often assumed to the independent. Our results provide correction to this approximation for the circular model. Another highlight is the modification of the amplitude of the joint Carpentier-Le Doussal (CLD) tail \cite{carpentier2001glass} by the max/min correlation (compared to product of marginals). We shall give a formula \eqref{eq:ratio} for the tail ratio for \textit{general} log-REM's.

\textit{Model and main results.} The \textit{circular model} of $1/f$-noise is defined as random signals $V_{j,M}, 1\leq j\leq M$, and their periodic extension on $\mathbb{Z}$ of period $M$, generated by independent Gaussian Fourier modes with variance proportional to inverse frequency:
\begin{equation}
V_{j,M} = \Re \left[ \sum_{k=-\frac{M}{2}}^{\frac{M}{2}-1} \sqrt{\frac{1}{|k|}} (u_k + \im v_k) \exp\left(\frac{ 2\pi \im kj}{M}\right) \right] \,. \label{eq:noisedef}
\end{equation}
Here $\{u_k, v_k\}_k$ are i.i.d standard centered Gaussian \cite{rosso12counting}. The definition implies a logarithmically growing variance
\begin{equation} \overline{V_{j,M}^2} = 2(\ln M + W) \, , \;  W \rightarrow \gamma_E - \ln 2 \label{eq:variance}
\end{equation}
characteristic of all (log)-REM's, and the off-diagonal correlations describing the planar GFF on the unit circle. Indeed, setting
\begin{equation}
\xi_{j,M} =  \exp\left(\frac{ 2\pi \im j}{M}\right)\, , \label{eq:xijm}
\end{equation} 
then for any sequence of pairs $(j_M, k_M)$ such that $\left(\xi_{j_M,M}, \xi_{k_M,M} \right) \rightarrow (\xi, \eta)$ with $\xi \neq \eta$, 
\begin{equation}
\overline{V_{j_M,M} V_{k_M,M} } \rightarrow 2\ln \abs{\xi - \eta} \,,\label{eq:gff}
\end{equation} 
which is the defining correlator of the planar GFF.

The observable $\xi_{j,M}$ in \eqref{eq:xijm} can be seen as a $O(2)$ (XY) spin (in the $M\rightarrow\infty$ limit). Its thermal average with inverse temperature $\beta > 0$ is:
\begin{align}
\langle \xi \rangle = \frac{\sum_{j=1}^M \exp(- \beta V_{j,M}) \xi_{j,M}}{\sum_{j=1}^M  \exp(- \beta V_{j,M}) } \, .\label{eq:andersondef}
\end{align}
We define the modulus square of the above as the EA order parameter of the circular model. Here we obtain its full temperature dependence:
\begin{equation}
\overline{\abs{\langle \xi \rangle}^2} \stackrel{M\rightarrow\infty}=
\begin{dcases} 
\frac{\beta^2}{1 + \beta^2} \, , & \beta \leq 1\, , \\
\frac{2\beta - 1}{2\beta}\, , & \beta \geq 1 \, .
\end{dcases} \label{eq:anderfreez}
\end{equation}

The \textit{minimum and maximum}, denoted as $V_{M\pm} = \pm \min_{j=1}^{M} \left( \pm V_{j,M} \right)$, 
are known for standard log-REM's to satisfy \cite{carpentier2001glass,bramson2012tightness}
\begin{align}
&V_{M\pm} = \mp 2\ln M \pm \frac{3}{2} \ln \ln M + v_{\pm} \pm c_M \, , \label{eq:Vleading}
\end{align}
where $c_M$ converges to some unknown deterministic constant as $M \rightarrow \infty$, while $v_{\pm}$ are the fluctuations. For the present circular model, $-v_-$ and $v_+$ have the same distribution \cite{fyodorov08rem}:
\begin{align}
&P(v_{+} > y) =2e^{\frac{y}{2}} K_1(2e^{\frac{y}{2}}) \Rightarrow \overline{\exp(t v_\pm)} = \Gamma^2(1 \pm t) \, ,\label{eq:vfb}
\end{align}
exhibiting the CLD tail \cite{carpentier2001glass} $P(v_{+}) \stackrel{v_+\rightarrow-\infty}\longrightarrow -v_+ e^{v_+}$ \footnote{$P(\text{an event})$ denotes its probability; $P(\text{random variable(s)})$ denotes their probability density function.}.
We generalise \eqref{eq:vfb} to the joint $v_{\pm}$ distribution:
\begin{align} 
&\overline{\exp(t_1 v_{+} - t_2 v_{-})} = S_{1}(t_1,t_2) \prod_{i=1}^2 \Gamma^{2}(1+t_i) \,, \label{eqetvv} \\ 
&S_{\beta}(t_1,t_2) = \sum_\lambda \prod_{\substack{(x,y)\in \lambda \\ i=1,2}} \frac{x \beta^{-1} + y \beta + t_i  }{(x + 1)\beta^{-1} + (y + 1)\beta + t_i}\, ,  \label{eqS2v}
\end{align}
where $S_{\beta}$ is a sum over partitions $\lambda = \{(x,y): x = 0, \dots, \lambda_y - 1 , y = 0 , \dots, l-1 \}$ (Here $\lambda_0 \geq  \dots \geq \lambda_{l-1} > 0$, $l \geq 0$ being the partition length; the empty partition is included). This implies the following joint/marginal   CLD tail ratio:
\begin{align}
R := \lim_{v_{\pm}\rightarrow\mp\infty}\frac{P(v_+,v_-)}{P(v_+)P(v_-)} = S_1 (-1,-1) = 2\, . \label{eq:carpentier}
\end{align} 
This result is a special case of the following ratio formula
\begin{equation}\label{eq:ratio}
R = \lim_{M\rightarrow\infty} \frac{1}{M^2}\sum_{j,k=1}^M \exp\left(- \beta_c^2 \overline{V_{j,M} V_{k,M}} \right)\, ,
\end{equation}
valid for\textit{ any} log-REM defined by the covariance matrix $\left[\, \overline{V_{j,M} V_{k,M}} \,\right]_{j,k=1}^M$ such that the above limit exists.

\textit{Joint $(v_{+}, v_{-})$ distribution.} Now we derive the joint distribution \eqref{eqetvv}; for this we study the thermodynamics at inverse temperature $\pm \beta$, encoded in the partition functions
\begin{equation}
\mathcal{Z}_{M\pm} =\sum_{j=1}^{M} \exp\left(\mp \beta V_{j,M}  \right) \, .
\end{equation}
When $\beta\rightarrow\infty$, the free energy $\mathcal{F}_{M\pm} := \mp \beta^{-1} \ln \mathcal{Z}_{M\pm} \rightarrow V_{M\pm}$. 
Define the regularised partition functions (\textit{c.f.} \eqref{eq:variance}) 
\begin{equation}\label{eq:norm_Z}
Z_{\pm} =  \mathcal{Z}_{M\pm} / \overline{\mathcal{Z}_{M\pm}} = \frac{\mathcal{Z}_{M\pm}}{M^{1 + \beta^2}e^{\beta^2 W}} \, . 
\end{equation}
Then, the replica averages $\overline{Z_+^mZ_-^n}$ converge to Coulomb--gas integrals as $M\rightarrow\infty$ if $\beta < \min\left(n^{-1/2}, m^{-1/2}\right)$ (in this work, we denote by $\stackrel{!}=$ equations that hold in the $M\rightarrow\infty$ limit and for $\beta$ small enough):
\begin{align}
&\overline{Z_+^mZ_-^n} \stackrel{!}{=}
 \int \mu_n^{\alpha}(\underline{\xi})\mu^{\alpha}_m(\underline{\eta}) \prod_{a,b} \abs{1 - \xi_a^* \eta_b}^{-2/\alpha}, \label{eqznmdef} \\
&1/\alpha = -\beta^2, \, \mu_n^{\alpha}(\underline{\xi}) = \prod_{a=1}^n \frac{\dif \xi_a}{2\pi\im \xi_a} \prod_{a<a'}  \abs{\xi_a-\xi_{a'}}^{2/\alpha}\, .  \label{eq:alphabeta} 
\end{align}
The integrals run on the unit circle $\abs{\xi_a} = \abs{\eta_b} = 1$ and the product runs from $a=1,\dots, n$ and $b = 1,\dots, m$. The notations introduced in \eqref{eq:alphabeta} is convenient for applying Jack polynomial theory to calculate the integral \eqref{eqznmdef}, the $m = n$ case of which appeared in \cite{saleur95kondo} for studying the Kondo problem. Their approach consists of two steps that we adapt to the present case, following conventions of \cite{macdonald1998symmetric} sect 6.10. First, one uses the Cauchy identity (\textit{ibid.}, (6.10.4) combined with 2nd paragraph of pp 380)
\begin{equation}\label{eqcauchy}
\prod_{a,b} (1 - \xi_a \eta_b)^{-1/\alpha} = \sum_\lambda P_\lambda^{(\alpha)}(\underline{\xi}) Q_\lambda^{(\alpha)}(\underline{\eta}) \,,
\end{equation}
where $P_\lambda^{(\alpha)}(\underline{\xi}) $ and $Q_\lambda^{(\alpha)}(\underline{\xi})$ form dual bases of Jack polynomials. Using  \eqref{eqcauchy} we expand the product in \eqref{eqznmdef}:
\begin{align}
&\prod_{a,b} \abs{1 - \xi_a^* \eta_b}^{-2/\alpha} = \prod_{a,b}  (1 - \xi_a \eta_b^*)^{-1/\alpha} \times (\textit{c.c.}) \nonumber \\
=& \sum_{\lambda, \mu} P_\lambda^{(\alpha)}(\underline{\xi}) Q_\lambda^{(\alpha)}(\underline{\eta^*})
 P_\mu^{(\alpha)}(\underline{\eta}) Q_\mu^{(\alpha)}(\underline{\xi^*}) \, .  \label{eq:denominator}
\end{align}
Then we apply the orthogonality relation (\textit{ibid.}, 6.10.35 -- 6.10.37)
\begin{align}
&\int \mu_n^{\alpha}(\underline{\xi}) P_\lambda^{(\alpha)}(\underline{\xi}) Q_\mu^{(\alpha)}(\underline{\xi^*}) = \delta_{\lambda \mu} p_n^\lambda(\alpha) c_n(\alpha) \, , \\
&c_n(\alpha) = \int \mu_n^{\alpha}(\underline{\xi})  = \frac{\Gamma(1 + n/\alpha)}{\Gamma(1 + 1/\alpha)^n} \, , \label{eqdyson} \\
& p_n^\lambda(\alpha) =   \prod_{(x,y)\in\lambda} \frac{\alpha x + n - y}{\alpha(x + 1) + n - (y + 1)}\, , \label{eqg}
\end{align}
where $c_n(\alpha)$ is the Dyson's integral \cite{dyson1962statistical1}. Orthogonality, combined with eqs. \eqref{eq:denominator}  and \eqref{eqznmdef}, yields the following:
\begin{align}
\overline{Z_+^n Z_-^m} \stackrel{!}= \frac{\Gamma(1 - n \beta^2)\Gamma(1 - m \beta^2)}{\Gamma(1 - \beta^2)^{m+n}}	 \sum_\lambda p_n^\lambda(\alpha) p_m^\lambda (\alpha) \,. \label{eqsmallzn}
\end{align}
In eqs. \eqref{eqdyson} through \eqref{eqsmallzn}, $n$ is continued to complex variable. The denominator in \eqref{eqsmallzn} can be absorbed by a first moment shift of the free energy:
\begin{align}
f_\pm :=  F_{\pm} \pm \frac{1}{\beta} \ln\Gamma(1 - \beta^2) \,,\,  F_{\pm} := - \beta^{-1} \ln Z_{\pm} \, . \label{eq:defsmallf} 
\end{align}
Now setting $t_1 = - n \beta, t_2 = - m \beta$ and using \eqref{eq:alphabeta}, \eqref{eqsmallzn} is rewritten as
\begin{align}
\overline{\exp({t_1 f_+  - t_2 f_- })} \stackrel{!}{=}  S_{\beta}(t_1,t_2)\prod_{i=1}^{2} \Gamma(1+\beta t_i) \,,\label{eqetfpm}
\end{align}
where $S_\beta(t_1, t_2)$ is given by \eqref{eqS2v}. Eq. \eqref{eqetfpm} holds actually for any $\beta < 1$ and generic complex $t$: this claim relies on assuming analyticity in the $\beta < 1$ phase and is non-rigorous (however, see \cite{ostrovsky2009mellin}), but can be numerically checked with high precision. Now, we observe several familiar features: the $\mp(\beta + \beta^{-1})\ln M$ extensive free energy, the UV-divergences of the integral \eqref{eqznmdef}, and the divergence of $\Gamma(1-\beta^2)$
\footnote{This divergence was argued to be the precursor of the $\frac{3}{2}\ln\ln M$ correction in \eqref{eq:Vleading}. A first-principle demonstration of this point is still missing; however, see \cite{rosso12counting} sect. 3 and \cite{fyodorov2015high} sect. 1.}, suggesting that \eqref{eqetfpm} is valid only until $\beta = \beta_c = 1$ and the $\beta > 1$ (glassy) phase should be described by the FDC \eqref{eq:freezdual}. 
The latter can be applied because the RHS \eqref{eqS2v} is duality-invariant: transpose partition pairs give terms related by the duality transform $\beta \leftrightarrow \beta^{-1}$. Indeed, introducing two i.i.d. standard Gumbel variables $g_\pm$ jointly independent of $f_{\pm}$, and constructing the usual duality-invariant\textit{ decoration} of the free energy, here at inverse temperature $\pm \beta^{-1}$: 
$ y_{\pm\beta}:= f_\pm \mp \beta^{-1} g_\pm$, we have $\overline{\exp(t_1 y_{+\beta} - t_2 y_{-\beta})} \stackrel{\beta<1}= S_\beta(t_1,t_2) \prod_{i=1}^{2}\Gamma(1 + \beta t_i) \Gamma(1 + t_i / \beta)$.
From the FDC \eqref{eq:freezdual}, the duality-invariance of the RHS implies the freezing of the LHS:
\begin{equation}\label{eq:lowTff}
\overline{\exp(t_1 f_{+} - t_2 f_{-})} \stackrel{\beta>1}= S_1(t_1,t_2) \prod_{i=1}^{2}\frac{\Gamma^2(1 + t_i)}{\Gamma(1 + t_i/\beta)}\, ,
\end{equation}
which yields \eqref{eqetvv} when $\beta\rightarrow\infty$. The novelty here is the extension of the FDC to the \textit{joint} distribution of opposite extrema. 
 
A nice extension of the above result is as follows. Let $q \in (0,1)$, consider two circular models $V_{j,M}^{(\pm)},  j = 1,\dots, M$, \textit{correlated} as $\overline{V_{j,M}^{(+)}V_{k,M}^{(-)}} = -2\ln \abs{1 - q \xi_{j,M} \xi_{k,M}^*}$, \footnote{One may see this as placing two circles at radii $1$ and $q$ in $\C$ endowed with \textit{one} GFF with correlator \eqref{eq:gff}.} and let $f_\pm$ \eqref{eq:defsmallf} be defined with respect to $V_{j,M}^{\pm}$. Then a direct extension of the above derivation leads to
\begin{align} 
& \frac{\overline{\exp(t_1 f_{+} - t_2 f_{-})}}{\overline{\exp(t_1 f_{+})}\times\overline{\exp(- t_2 f_{-})}} = S_{\min(\beta,1)}^{(q)}(t_1,t_2)\, , \label{eq:extensionq}\\ 
&S_{\beta}^{(q)}(t_1,t_2) = \sum_\lambda \prod_{\substack{(x,y)\in \lambda \\ i=1,2}} \frac{q\left( x \beta^{-1} + y \beta + t_i \right) }{(x + 1)\beta^{-1} + (y + 1)\beta + t_i}\, , \nonumber 
\end{align}
interpolating between two independent circular models ($q\rightarrow0$) and \eqref{eqetvv} ($q\rightarrow1$). 

Now, to obtain the joint CLD tail behaviour at $v_{\pm}\rightarrow\mp \infty$, observe that the rightmost pole of \eqref{eqetvv} is $\frac{S_1^{(q)}(-1,-1)}{(1 + t_1)^2(1+t_2)^{2}}$, and that only $\lambda=\emptyset$ and $\square$ contribute to the sum $S_1^{(q)}(-1,-1) = 1 + q^2$. Therefore
\begin{align}
& \text{pdf}(v_{\pm}\rightarrow \mp \infty) \simeq  - (1+q^2) v_+ e^{v_+} v_- e^{-v_-} \, , \label{eq:carpentier1}
\end{align}
which reduces to \eqref{eq:carpentier} for $q = 1$. In fact, the value $R = S_{1}(-1,-1) = 2$ can be explained as follows. Tracking back the derivation, we can see that 
$ S_\beta(-n\beta,-m\beta) \stackrel{!}= \frac{\overline{Z_+^n Z_-^m}}{\overline{Z_+^n}\,\overline{Z_-^m}}.$
Thus $-n\beta = -m\beta = -1, \beta = 1 \Rightarrow m = n = 1$, so by definition of $Z_{\pm}$, 
\begin{equation}
R = S_{1}(-1,-1) = \lim_{M\rightarrow\infty} \frac{1}{M^2}\sum_{j,k} \exp(- \overline{V_{j,M} V_{k,M}})\, ,
\end{equation}
recovering \eqref{eq:ratio} for the circular model (that $\text{RHS}=2$ is elementary). Both sides of \eqref{eq:ratio} are defined for general log-REM's; moreover, its derivation here is \textit{in fine} independent of the circular model context. Thus we conjecture that the relation \eqref{eq:ratio} holds \textit{for generic log-REM's}. 

\textit{Application to diffusion.} Consider a particle hopping in 1d infinite lattice, driven by a periodic potential $V_j \equiv V_{j \text{ mod } M,M}$. Let the dynamics be continuous-time Markov nearest neighbour hoppings, whose rates $W_{j\rightarrow i}$ are function of $V_i$ and $V_j$ satisfying detailed balance. The long time dynamics is diffusive $\langle (j_t - j_0)^2 \rangle \simeq  D_V t$; we rescale the time so that when $V \equiv 0$, the diffusion constant $D_V = 1$. Using results of \cite{derrida1983velocity}, one can show $D_V = M^2 / \left(\mathcal{Z}_+ \mathcal{Z}_-\right)$. \footnote{The relation holds for $M\rightarrow +\infty$ (and for the Langevin
continuum version, see \textit{e.g.} \cite{dean14diffusion,ledoussal1995creep}). Note that $D_V$ is disorder-dependent, not to be confused with disorder-averaged definitions see \textit{e.g.} \cite{ledoussal1989annealed}. } Its typical value is
\begin{equation} D_{\text{typ}} = \begin{cases}
 \Gamma^2(1-\beta^2) M^{-2\beta^2}\, , & \beta < 1\, , \\  a_1 \ln (M/b_1) M^{-2}\, , & \beta = 1\, , \\ a_2 \ln^{3\beta}(M/b_2) M^{-4\beta+2}\, , & \beta> 1\, .
\end{cases}\end{equation}
where $a_{1,2}, b_{1,2}$ are unknown constants. Now, eqs. \eqref{eqetfpm} and \eqref{eq:lowTff} describe the fluctuation of $D_V$ around $D_{\text{typ}}$ in terms of the Mellin transform
\begin{equation}
\overline{\left(\frac{D_V}{D_{\text{typ}}}\right)^s} \stackrel{M\rightarrow\infty}= \begin{dcases}   \Gamma^2(1 + s \beta^2) S_{\beta}(s\beta,s\beta)\, ,& \beta\leq 1\, , \\
 \frac{\Gamma^4(1 + s\beta)}{\Gamma^2(1 + s)} S_1(s\beta,s\beta)\, , & \beta > 1\, .
\end{dcases}
\end{equation}
We remark that a closely related dynamical quantity is the sum of left and right \textit{first-passage} times. Placing the particle at $0$ at $t = 0$, consider 
$\tau_\pm = \min \{ t: j_t = \pm M \}$, then their thermal average satisfies \cite{castillo2001freezing} $\langle \tau_+ + \tau_- \rangle =  D_V^{-1} M^2,$ to which the above statements apply.

\begin{figure}
\center\includegraphics[scale=0.35]{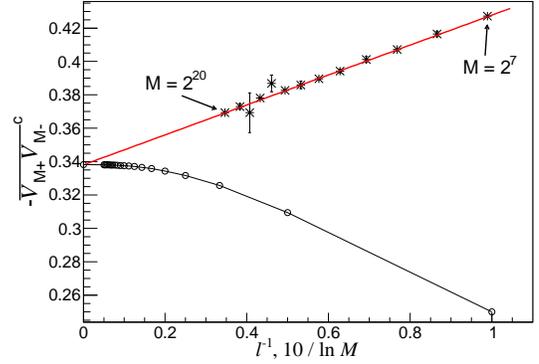}
\caption{Numerical check of the (minus) min-max covariance $ -\overline{V_{M+} V_{M-}}^{c}$. The $1/f$--noise \eqref{eq:noisedef} is generated using Fast Fourier transform, with $\geq 10^6$ independent realisations for each size. The numerical data ($*$) $2^7 \leq M \leq 2^{20}$ are consistent finite-size scaling $a + b / \ln M$, with $b = 0.89(1)$ and $a = 0.338(1)$, in $3$-digit agreement with \eqref{eqsumcov}. The sums over partitions in this work are all convergent and calculated by the method of \cite{cao15gff}, which involves a truncation size $l$. The sum \eqref{eqsumcov} truncated to $l = 1, \dots, 20$ are ploted ($\circ$) to appreciate convergence; in all cases $l \sim 10^2$ yields sufficient precision. } \label{figcov}
\end{figure}

\textit{Numerics on max-min correlation}. The first consequence of eqs. \eqref{eq:Vleading} and \eqref{eqetvv} is the min--max covariance:
\begin{align}
& -\overline{V_{M+} V_{M-}}^{c} \stackrel{M\rightarrow\infty}\longrightarrow  -\overline{v_{+} v_{-}}^{c} = \left. \frac{\partial^2 S_{1}}{\partial t_1 \partial t_2}\right\vert_{t_1,t_2 = 0} \nonumber \\
 = & \sum_{\lambda\neq \emptyset} 
 \frac{1}{4} \prod_{x,y \neq (0,0)} \frac{(x + y)^2}{(x + y + 2)^2}  = 0.338\dots \, , \label{eqsumcov}
\end{align}
which is a rather small correlation compared with $\overline{v_{\min}^2}^{c} = \overline{v_{\max}^2}^{c} = \pi^2 / 3 = 3.29\dots$. Yet its persistence in the $M\rightarrow\infty$ limit is strongly supported by our numerical analysis, see fig. \ref{figcov}. Remark that to achieve quantitative agreement at zero temperature the finite-size effects must be accounted for. Here we find this can be done by a linear form in $(\ln M)^{-1}$.

An heuristic explanation of the negative $v_{\pm}$ correlation is that, every term in \eqref{eq:noisedef} is a plane wave that pushes $V_{M\pm}$ to opposite directions. 
For comparison, in the case of the Cayley tree (or branching Brownian motion, BBM) model \cite{derrida1988polymers}, the $v_{\pm}$ correlation is \textit{positive}, since it originates from their common ancestor. In both BBM and circular model, although persisting in the thermodynamic limit, the correlation is vanishingly \textit{weak} compared to $V_{M\pm}$. In this respect, let us mention the \textit{strong} min--max correlation exhibited by Ramola \textit{et. al.} \cite{ramola2015spatial} in a generalised BBM with particles dying and splitting at tunable rates. We are not aware of any non-hierarchical analogue. 

\begin{figure}[h]
\center \includegraphics[scale=0.35]{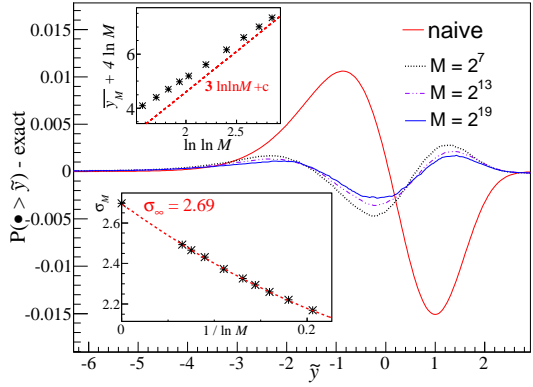} 
\caption{Main: The cumulative distribution of the span $y_M$ rescaled to mean$=0$ and variance$=1$, from which we subtract the exact prediction \eqref{eqityinfty}. The red line is the naïve prediction \eqref{eqityguess}. Both predictions are obtained by inverse Fourier transform; the partition sum is calculated by the method of \cite{cao15gff} with $l = 200$ truncation. Top inset: the next-to-leading behaviour of the span $\overline{y_M} + 4 \ln M$, compared to the prediction $3 \ln\ln M + c$ of \eqref{eqityinfty}. Bottom inset: Finite-size scaling of span standard variation $\sigma_M$. The dots are numerical data, the red line represents the Ansatz $a_0 + a_1 (\ln M)^{-1} + a_2 (\ln M)^{-2}$, with $a_0 = 2.69(1), a_1 = - 3.29(3)$ and $a_2 = 3.5(1)$. } \label{figcdfvar}
\end{figure}

As another numerical check, we consider the \textit{span}, \textit{i.e.} the difference the two extrema. 
Its distribution is inferred from eqs. \eqref{eq:Vleading} and \eqref{eqetvv}:
\begin{align} & y_M := V_{M+} - V_{M-} = -4\ln M + 3\ln\ln M +  y + 2c_M \,, \nonumber \\ 
& \overline{\exp(t y)} = \Gamma^4(1 + t) S_1(t,t). \label{eqityinfty}
\end{align} 
The naïve approximation can be obtained by discarding the non-trivial sum $S_1$ encoding the $v_{\pm}$ correlation:
\begin{equation}
 \overline{\exp(t y)} \simeq \Gamma^4(1 + t)\,. \label{eqityguess}
\end{equation}
Now we compare predictions \eqref{eqityinfty} and \eqref{eqityguess} against numerical measures of $y_M$.  We consider separately the variance $ \sigma_M^2 = \overline{y_M^2}^c$ and the rescaled distribution $\tilde{y}_M = (y_M - \overline{y_M}) / \sigma_M$. As shown in fig. \ref{figcdfvar} main, the numerical cumulative distribution converges to the exact prediction \eqref{eqityinfty} and rules out the naïve prediction eqs. \eqref{eqityguess}. More convincing evidence can be obtained by considering the variance, by taking into account the finite-size correction (fig. \ref{figcdfvar} bottom inset). Analytically, \eqref{eqityinfty}, \eqref{eqsumcov} and \eqref{eq:vfb} imply $ \sigma_{M} \rightarrow 2.6937\dots$, in fine agreement with the numerical value $\sigma_{\infty} = 2.69(1)$ obtained by a quadratic finite-size Ansatz. 

In both cases, simple finite-size Ansätze improve the quality of numerical evidence of freezing by giving very good agreements at zero temperature. Yet, it is still hard to check conclusively the CLD tail ratio \eqref{eq:carpentier}, or its consequence on the span \footnote{We also obtain that $P(y\rightarrow-\infty)\simeq - 2 \frac{y^3}{6} e^{y}$ from \eqref{eqityinfty}, while the approximation \eqref{eqityguess} would miss the factor $2$.}, which requires both large $M$ and high statistics. Nonetheless, we hope the more general formula \eqref{eq:ratio} be tested in physical or numerical experiments.

\textit{Edwards-Anderson's order parameter.} Sums over partitions here and elsewhere \cite{fyodorov2015moments,cao15gff} share the duality-invariance structure, providing an infinite series of freezing observables, yet to be interpreted. Here we show that, the first term of the sum in \cite{cao15gff} encodes the Edwards-Anderson's (EA) order parameter \eqref{eq:andersondef}, whose glassy phase behaviour is a non-trivial consequence of the FDC.

Let us recall the partition sum studied in \cite{cao15gff}:
\begin{equation}\label{eq:sum}
c_n^{-1}(\alpha)\int \mu_n^\alpha(\underline{\xi}) \prod_{a,b}(1-q \xi_a \xi_b^*)^{-\frac{1}{\alpha}}
 =   \sum_{\lambda} q^{|\lambda|}  p_n^\lambda(\alpha) \,,
\end{equation}
where $\mu, c$ and $p$ are defined in eqs. \eqref{eq:alphabeta},\eqref{eqdyson} and \eqref{eqg}, $q \in (-1, 1)$, and $\abs{\lambda}$ is the size of the partition. Viewing both sides as power series of $q$, at order $q^1$, we have (\textit{c.f.} \eqref{eqg})
\begin{equation}
\frac{1}{\alpha c_n(\alpha)} \int \mu_n^\alpha(\underline{\xi}) \sum_{a,b} \xi_a \xi_b^{*} = p_n^{\square}(\alpha) =  \frac{ n }{n - 1 + \alpha}\, ,  \label{eq:qterm}
\end{equation}
which, upon the usual change of variables, is duality-invariant:
\begin{equation} n = -t/\beta, \, 1/\alpha = -\beta^2 \Rightarrow  p_n^{\square}(\alpha) = \frac{t}{t+ \beta + \beta^{-1}} \,. 
\label{eq:pnsquare}\end{equation}
But in order to apply the FDC we need to interpret the LHS as the $M\rightarrow\infty$ expression of some observable in the $\beta < 1$ phase. For this, we calculate
\begin{align}
\int \mu_n^\alpha(\underline{\xi}) \sum_{a,b} \xi_a \xi_b^{*}& = n c_\alpha(n) +  n(n-1) \int  \mu_n^\alpha(\underline{\xi}) \xi_1 \xi_2^* \nonumber \\
\stackrel{!}=& n \overline{Z_+^n} +  n(n-1)  \overline{Z_+^n \langle\xi\rangle \langle \xi^* \rangle  } \nonumber \\
  = & - \frac{t}{\beta^2} \overline{e^{tF_+} \left( \beta - (t + \beta) \abs{\langle \xi \rangle}^2  \right)} \, . \label{eq:replica_anderson}
\end{align}
The first equality exploits the permutation symmetry of $\mu_n^\alpha(\underline{\xi})$ and applies the definition of Dyson integral; the second one is a standard replica calculation (similar to \eqref{eqznmdef}) using \eqref{eq:andersondef}. Combining \eqref{eq:qterm} through \eqref{eq:replica_anderson} and \eqref{eqg} gives
\begin{align}
\overline{e^{t f_+}\left(\beta - (t + \beta) \abs{\langle \xi \rangle}^2 \right)} \stackrel{\beta<1}{=} \frac{\Gamma(1+t\beta)}{t + \beta + \beta^{-1}}\, , \label{eq:etfandersonhighT}
\end{align}
The usual decoration $y_\beta := f_+ - \beta^{-1} g$ where $g$ is an independent Gumbel is again valid here: indeed \eqref{eq:etfandersonhighT} implies
\begin{equation} \overline{e^{ty_\beta}\left(\beta - (t + \beta) \abs{\langle \xi \rangle}^2 \right)} \stackrel{\beta<1}{=} \frac{\Gamma(1 + t\beta)\Gamma(1 + t/\beta)}{t + \beta + \beta^{-1}}\,. 
\end{equation} 
The duality-invariance of RHS triggers the freezing of the LHS, yielding
\begin{equation}\label{eq:etfandersonlowT}
\overline{e^{t f_+}\left(\beta - (t + \beta) \abs{\langle \xi \rangle}^2 \right)}  \stackrel{\beta > 1}{=} \frac{\overline{e^{t f_+}}}{t + 2}\, .
\end{equation} 
Equations \eqref{eq:etfandersonhighT} and \eqref{eq:etfandersonlowT} can be summarised as
\begin{align}\label{eq:anderson_general}
\left. \overline{e^{tf_+}\abs{\langle \xi \rangle}^2 \rangle} \right/  \overline{e^{tf_+}} =
\begin{dcases}
   \frac{\beta^2}{\beta^2 + t\beta + 1}\,, & \beta \leq 1\,,  \\
   \frac{(t+2)\beta-1}{(t + \beta)(t+2)}\,, & \beta > 1\,,
 \end{dcases}
\end{align}
which, as a series in $t$, generates the joint-cumulants of $f_+^{k}$ and $\abs{\langle \xi \rangle}^2$:
\begin{align}
 &\sum_{k = 0}^{\infty} \frac{t^k}{k!} \overline{f_+^{k} \abs{\langle \xi \rangle}^2}^{c} = \left. \overline{e^{tf_+}\abs{\langle \xi \rangle}^2} \right/  \overline{e^{tf_+}}\, , \\
&\overline{ \frac{(- f_+)^{k}}{k!} \abs{\langle \xi \rangle}^2 }^{c} = 
\begin{dcases}
\frac{\beta^{k+2}}{(\beta^2 + 1)^{k+1}}\,, & \beta \leq 1 \,, \\
\frac{1}{\beta^k} + \frac{\frac{1}{\beta^{k+1}} -\frac{1}{2^{k+1}} }{\beta - 2} \,,   & \beta > 1 \, .  
\end{dcases}\label{eq:andersonf_moment}
\end{align}
At order $k = 0$, we retrieve \eqref{eq:anderfreez}. The last prediction is conclusively confirmed by the numerics, see fig. \ref{fig:anderson}. 

Using an inverse Laplace/Fourier transform on \eqref{eq:anderson_general} we obtain $\overline{ \abs{\langle\xi\rangle}^2 }^f$, the EA order parameter {\it conditioned on the free energy $f_+=f$}. For the $\beta < 1$ case, we have 
\begin{equation}
(\beta^2 + 1 - \beta \partial_f )  \overline{\delta(f_+ - f) \abs{\langle\xi\rangle}^2}\stackrel{\beta<1}=  \beta^2 \overline{\delta(f_+ - f)} \,,
\end{equation}
where $\overline{\delta(f_+ - f)} = \beta^{-1} \exp(f/\beta - e^{f/\beta})$ \cite{fyodorov08rem}. Requiring $ \overline{ 0 \leq \abs{\langle\xi\rangle}^2 }^f \geq 1$, a unique solution is obtained:
\begin{equation}
 \overline{ \abs{\langle\xi\rangle}^2 }^f
 = \frac{\overline{\abs{\langle\xi\rangle}^2\delta(f_+ - f)}}{\overline{\delta(f_+ - f)}} \stackrel{\beta<1}=  1-e^{\beta f+e^{f/\beta}} \Gamma(1-\beta^2,e^{f/\beta}) \label{eq:conditionedEA}
\end{equation}
where $\Gamma(s,z) = \int_z^{+\infty} e^{-x} x^{s-1} \dif x$ is the incomplete Gamma function. This yields the
asymptotics
$\overline{ \abs{\langle\xi\rangle}^2 }^f \stackrel{f \to -\infty}{\simeq} 1- e^{\beta f} \Gamma(1-\beta^2)$
and 
$\overline{ \abs{\langle\xi\rangle}^2 }^f \stackrel{f \to +\infty}{\simeq} \beta^2 e^{-f/\beta}$.
The $\beta > 1$ phase calculation follows the same principle and will be omitted for being more cumbersome.

\textit{Conclusion.}--
We calculated the joint min-max distribution and the Edwards-Anderson's order parameter of the circular $1 / f$-noise model, as well as generalisations. Each of them provides a numerically convincing test of the freezing-duality conjecture. Its implementations are variants of the usual decoration of the free energy distribution; it would be interesting to see how the mathematical treatment \cite{madaule2013glassy} can be adopted to cover these cases. The treatment on the EA order parameter is an example to be generalised to further terms, which provide an infinite series of duality-invariant observables, indexed by (pairs of) partitions, and hopefully a clarification on 	the origin and generality of the duality invariance. 

\acknowledgments
We thank Y. Fyodorov and A. Rosso for useful discussions. We acknowledge support from PSL grant ANR-10-IDEX-0001-02- PSL. We thank the hospitality of KITP, under Grant No. NSF PHY11-25915.

\bibliographystyle{eplbib.bst}
\bibliography{rems}
\end{document}